\documentclass[twocolumn]{aastex62}

\usepackage{epsfig}
\usepackage{epstopdf}
\usepackage{graphics}
\usepackage{amsmath}
\usepackage{xcolor}
\usepackage{aas_macros}

\usepackage{array}
\usepackage{booktabs}
\usepackage{amsmath,amssymb,amsfonts,amsbsy}
\usepackage{mathrsfs}
\usepackage{url}
\usepackage{multirow}
\usepackage{xcolor}
\usepackage{color}
\usepackage{ulem}
\usepackage{enumerate}

\renewcommand{\vec}[1]{\boldsymbol{#1}}

\newcommand{\dif}{\mathrm{d}}

\renewcommand{\figurename}{Fig.}
\renewcommand{\tablename}{Tab.}
\begin{document}

\title{Geminga SNR: Possible candidate of local cosmic-ray factory}

\correspondingauthor{Wei Liu, Xun-Xiu Zhou}
\email{liuwei@ihep.ac.cn, zhouxx@home.swjtu.edu.cn}

\author{Bing Zhao}
\affiliation{Southwest Jiaotong University, 610031 Chengdu, Sichuan, China
}

\author{Wei Liu}
\affiliation{Key Laboratory of Particle Astrophysics,
	Institute of High Energy Physics, Chinese Academy of Sciences, Beijing 100049, China
}

\author{Qiang Yuan}
\affiliation{Key Laboratory of Dark Matter and Space Astronomy,
Purple Mountain Observatory, Chinese Academy of Sciences,
Nanjing 210008, China
}
\affiliation{School of Astronomy and Space Science, University of Science and Technology of China,
Hefei 230026, China
}
\affiliation{Center for High Energy Physics, Peking University,
Beijing 100871, China}

\author{Hong-Bo Hu}
\affiliation{Key Laboratory of Particle Astrophysics,
	Institute of High Energy Physics, Chinese Academy of Sciences, Beijing 100049, China
}
\affiliation{University of Chinese Academy of Sciences, Beijing 100049, China
}

\author{Xiao-Jun Bi}
\affiliation{Key Laboratory of Particle Astrophysics,
	Institute of High Energy Physics, Chinese Academy of Sciences, Beijing 100049, China
}
\affiliation{University of Chinese Academy of Sciences, Beijing 100049, China
}

\author{Han-Rong Wu}
\affiliation{Key Laboratory of Particle Astrophysics,
	Institute of High Energy Physics, Chinese Academy of Sciences, Beijing 100049, China
}

\author{Xun-Xiu Zhou}
\affiliation{Southwest Jiaotong University, 610031 Chengdu, Sichuan, China
}

\author{Yi-Qing Guo}
\affiliation{Key Laboratory of Particle Astrophysics,
	Institute of High Energy Physics, Chinese Academy of Sciences, Beijing 100049, China
}
\affiliation{University of Chinese Academy of Sciences, Beijing 100049, China
}


\begin{abstract}
The precise measurements of energy spectra and anisotropy could help us uncover the local cosmic-ray accelerators. Our recent works have shown that spectral hardening above $200$ GeV in the energy spectra and transition of large-scale anisotropy at $\sim 100$ TeV are of local source origin. Less than $100$ TeV, both spectral hardening and anisotropy explicitly indicate the dominant contribution from nearby sources. In this work, we further investigate the parameter space of sources allowed by the observational energy spectra and anisotropy amplitude. To obtain the best-fit source parameters, a numerical package to compute the parameter posterior distributions based on Bayesian inference, which is applied to perform an elaborate scan of parameter space. We find that by combining the energy spectra and anisotropy data, the permissible range of location and age of local source is considerably reduced. When comparing with the current local SNR catalog, only Geminga SNR could be the proper candidate of the local cosmic-ray source.
\end{abstract}

\section{Introduction}


As far back as 1930s, the supernova remnants (SNRs) had been proposed as the sources of Galactic cosmic rays (CRs) \citep{1934CoMtW...3...79B}. However due to the diffusive character of CR propagation in the Galaxy, it is hardly to locate their acceleration sites by tracing back the arrival directions of CRs, with a possible exception of ultra-high energy CRs. What have been achieved in identifying CR sources up to now are mostly indirect, namely through the multi-wavelength observations of electromagnetic emission from SNRs, see the reference of \citep{2014NuPhS.256...65C, 2015ARNPS..65..245F, 2015A&ARv..23....3D, 2015CRPhy..16..674H}. On the other hand, \cite{2011JCAP...02..031M, 2012A&A...544A..92B} have shown that CRs from the nearby young sources, within $1 - 2$ kpc from the solar system, could give rise to large fluctuations of observed energy spectrum. As a result, the unexpected observational features probably relate to these local sources. Meanwhile within this short distance, the CR sources are finite and discrete, thus they are easier to be found by the multi-wavelength observations. Therefore it is promising to directly unveil the local CR accelerators by relating the observational features of CRs to the local source.



In fact, the single source model, in which one or a few nearby young sources make nonnegligible contribution to the spectrum, was initially put forward for the purpose of the sharpness of the knee region at $\sim 3-4$ PeV in the all-particle spectrum \citep{1997JPhG...23..979E}. Later with more and more advanced instruments being put into use, the measuring accuracy has been promoted greatly and more novel features in the energy spectrum are uncovered. The single source model and its extension - local source model, are widely used to interpret various observational phenomena. Usually, the propagation of CRs from the nearby source is time-dependent, and the propagated spectrum resembles a bump-like structure, which is deemed as an excess of CR flux. Thus the local pulsar or SNR could be the naturally origin of the positron excess above $10$ GeV \citep{2009Natur.458..607A, 2009PhRvL.102r1101A, 2014PhRvL.113l1101A, 2014PhRvL.113l1102A} and spectral hardening of nuclei above $200$ GeV \citep{2007BRASP..71..494P, 2009BRASP..73..564P, 2011ApJ...728..122Y, 2011Sci...332...69A, 2015PhRvL.114q1103A,
2015PhRvL.115u1101A, 2019PhRvL.122r1102A} and ensuing softening at $\sim 20$ TeV \citep{2017ApJ...839....5Y, 2017JCAP...07..020A, 2019SciA....5.3793A}. Meanwhile it could also account for the break in all-electron spectrum at TeV energies \citep{2009A&A...508..561A, 2011ICRC....6...47B, 2015arXiv151001269S, 2017Natur.552...63D}. Furthermore, in the traditional propagation model, the predicted anisotropy amplitude from background SNRs far exceeds the measurements, which is only about $10^{-4} - 10^{-3}$ \citep{2012JCAP...01..011B}. The local source could effectively lower the amplitude, if it lies close to the direction of anti-Galatic center \citep{2016PhRvL.117o1103A, 2017PhRvD..96b3006L}.

In recent works, we established a coherent picture to explain both observed spectral features and anisotropy \citep{2019JCAP...10..010L, 2019JCAP...12..007Q}. We find that the amplitude transition and phase flipping in the dipole anisotropy map have a common origin with the spectral hardening of nuclei above $200$ GeV and ensuing falloff at $\sim 20$ TeV. Less than $100$ TeV, the anisotropy and spectral features are dominated by the local source. The position of local source is close to the direction of anti-Galatic center and far from the Galactic disk. We find that the Geminga SNR at its pulsar's birth place could be a prime candidate.


In fact, Geminga pulsar has long been considered as a local positron source, since the discovery of the increasing positron fraction above $10$ GeV. Recently HAWC experiment measured the extended TeV gamma-ray emission of Geminga and PSR B0656+14 pulsars \citep{2017Sci...358..911A}. The inferred diffusion coefficient nearby the $\gamma$-ray emission region is far less than the standard value derived by fitting the B/C ratio. It suggested the positron excess may have an exotic origin. However \cite{2018ApJ...863...30F} and \cite{2019MNRAS.484.3491T} argued that even the inference in view of the HAWC surface brightness profile is correct, the positron excess could still be accounted for by the Geminga pulsar, as long as introducing a two-zone diffusion model.

In this work, we aim at the parameter space of cosmic-ray sources permitted by the observed energy spectra and anisotropy. To perform the elaborate scan of parameter space of sources, the multinest package, based on Bayesian inference, is applied. By fitting the energy spectra and anisotropy amplitude, the permissible space of location and age of local source is greatly reduced. Our study further demonstrates that the Geminga SNR could be the best candidate of local cosmic-ray source.

The rest paper is organized as follows: In Sec.2, the propagation model and Bayesian inference are briefly introduced. Sec.3 presents the calculated results and Sec.4 is reserved for the conclusion.

\section{Model Description}

\subsection{Propagation Model}
The spatial-dependent propagation (SDP) model has received a lot of attention in recent years. It was first introduced as a Two Halo model (THM) \citep{2012ApJ...752L..13T} to explain the spectral hardening of both proton and helium  above $200$ GeV\citep{2011Sci...332...69A}. Afterwards, it is further applied to secondary and heavier components \citep{2015PhRvD..92h1301T, 2016PhRvD..94l3007F, 2016ApJ...819...54G, 2018ApJ...869..176L, 2020ChPhC..44h5102T, 2020arXiv200701768Y}, diffuse gamma-ray distribution \citep{2018PhRvD..97f3008G} and large-scale anisotropy \citep{2019JCAP...10..010L, 2019JCAP...12..007Q}. For a comprehensive introduction, one can refer to \cite{2016ApJ...819...54G} and \cite{2018ApJ...869..176L}.

In the SDP model, the whole diffusive halo is divided into two parts. The Galactic disk and its surrounding area are called the inner halo (IH) region, in which the diffusion coefficient is spatial dependent and relevant to the radial distribution of background CR sources. The extensive diffusive region outside the IH is named as the outer halo (OH) region, where the diffusion is regarded as only rigidity dependent. The size of IH is represented by its half thickness $\xi z_{h}$, whereas the OH region's is $(1-\xi) z_{h}$. The diffusion coefficient $D_{xx}$ in the diffusive halo is thus parameterized as:
\begin{equation}
D_{xx}(r,z, {\cal R} )= D_{0}F(r,z)\beta^{\eta} \left(\dfrac{\cal R}{{\cal R}_{0}} \right)^{\delta_0 F(r,z)} ~.
\label{eq:diffusion}
\end{equation}
where
\begin{equation}
 F(r,z) =
\begin{cases}
g(r,z) +\left[1-g(r,z) \right] \left(\dfrac{z}{\xi z_0} \right)^{n} , &  |z| \leq \xi z_0 \\
1 ~, & |z| > \xi z_0
\end{cases},
\end{equation}
with $g(r,z) = N_m/[1+f(r,z)]$. $f(r,z)$ is the source density distribution, which is approximated as axisymmetric, i.e. $f(r, z) = \left(\dfrac{r}{r_\odot} \right)^\alpha \exp \left[-\dfrac{\beta(r-r_\odot)}{r_\odot} \right] \exp \left(-\dfrac{|z|}{z_s} \right)$
with $r_\odot \equiv 8.5$ kpc and $z_{s} = 0.1$  kpc. The parameters $\alpha$ and $\beta$ are taken as $1.09$ and $3.87$ respectively in this work \citep{2015MNRAS.454.1517G}. The propagation of CRs from local point source is time-dependent. As for the instantaneous injection, the spatial distribution is $\psi({\cal R}, \vec{r}, t) = Q({\cal R}) e^{- (\vec{r} - \vec{r}^\prime)^2/(4 D_{xx} t)}/(4\pi D_{xx} t)^{3/2}$. The energy spectra at sources are assumed to have a power-law of rigidity plus an exponential cutoff,
\begin{equation}
q({\cal R}) \propto {\cal R}^{-\nu} \exp \left(-\dfrac{{\cal R}}{{\cal R}_{\rm c} } \right) ~.
\end{equation}

In this work, the diffusion-reacceleration (DR) propagation model is adopted. The numerical package DRAGON \citep{2008JCAP...10..018E} is used to solve the diffusion equation to obtain the CR distribution. Less than tens of GeV, the CR fluxes are impacted by solar wind. The force-field approximation is applied to describe the solar modulation effect \citep{1968ApJ...154.1011G, 1987A&A...184..119P}.

\subsection{Bayesian Parameter Inference}


Bayesian analysis method has now been widely used in astrophysics, cosmology as well as particle physics for the parameter inference. In this work, to quantitatively evaluate the probability density distribution (PDF) of model parameters permitted by the observed CR spectra and anisotropy amplitude, the Bayesian inference approach is applied. Here we give a brief introduction to the Bayesian inference of parameters. For a more comprehensive review, one can refer to \cite{2008ConPh..49...71T, 2009A&A...497..991P, 2011ApJ...729..106T, 2016PhRvD..94l3007F}.


Bayesian inference is essentially to make use of prior PDF for the parameters of interest and the likelihood function supplied by the data to evaluate the posterior PDFs. Given the parameter set $\vec{\Theta}$ and the observational data $\vec{D}$, Bayes Theorem reads
\begin{align}
P(\vec{\Theta}|\vec{D}) = \dfrac{P(\vec{D}| \vec{\Theta}) P(\vec{\Theta})}{P(\vec{D})} ~.
\label{eq:bayes}
\end{align}
$P(\vec{\Theta}|\vec{D})$ is the posterior PDF of the parameters, while $P(\vec{\Theta})$ is the prior PDF of parameters before the observations are considered. $P(\vec{D}|\vec{\Theta}) = L(\vec{\Theta})$ is called the likelihood function, a function of $\vec{\Theta}$ given the data set $\vec{D}$. The likelihood function is defined as
\begin{align}
{\cal L}(\vec{\Theta}) = \exp \left(-\dfrac{1}{2} \chi^2(\vec{\Theta}) \right) ~,
\end{align}
with $\chi^2(\vec{\Theta})$ built from the data and model, i.e.
\begin{align}
\chi^2(\vec{\Theta}) = \sum_{i = 1}^{N_{\rm data}} \left(\dfrac{y_i^{\rm exp} -y_i^{\rm th}(\vec{\Theta}) }{\sigma_i} \right)^2 ~,
\end{align}
in which $y^{\rm exp}_i$ and $\sigma_i$ are the measured value and standard deviation in experiment, and $y_i^{\rm th}$ is the theoretical expectation under certain parameter set $\vec{\Theta}$.
The denominator in Eq. (\ref{eq:bayes}), named the Bayesian evidence, is obtained by computing the average of the likelihood under the prior, i.e.
\begin{align}
P(\vec{D}) = \int P(\vec{D}|\vec{\Theta}) P(\vec{\Theta}) \dif \vec{\Theta} ~.
\end{align}

In the parameter estimation, the inferences are obtained by taking samples from the posterior using MCMC sampling methods, where at equilibrium the chain contains a set of samples from the parameter space distributed according to the posterior. With the posterior samples, the marginal posterior PDFs and other estimations could be available straightforward.

In this work, we adopt the public MultiNest package \citep{2008MNRAS.384..449F, 2009MNRAS.398.1601F, 2019OJAp....2E..10F}, which implements the nested sampling algorithm \citep{2004AIPC..735..395S}. Compared with the traditional Markov chain Monte Carlo (MCMC) methods, the nested sampling could navigate the parameter space with complex, multi-modal posterior distribution until a well-defined termination point with high efficiency.

\section{Results}

\begin{figure*}
	\centering
	\includegraphics[width=0.45\textwidth]{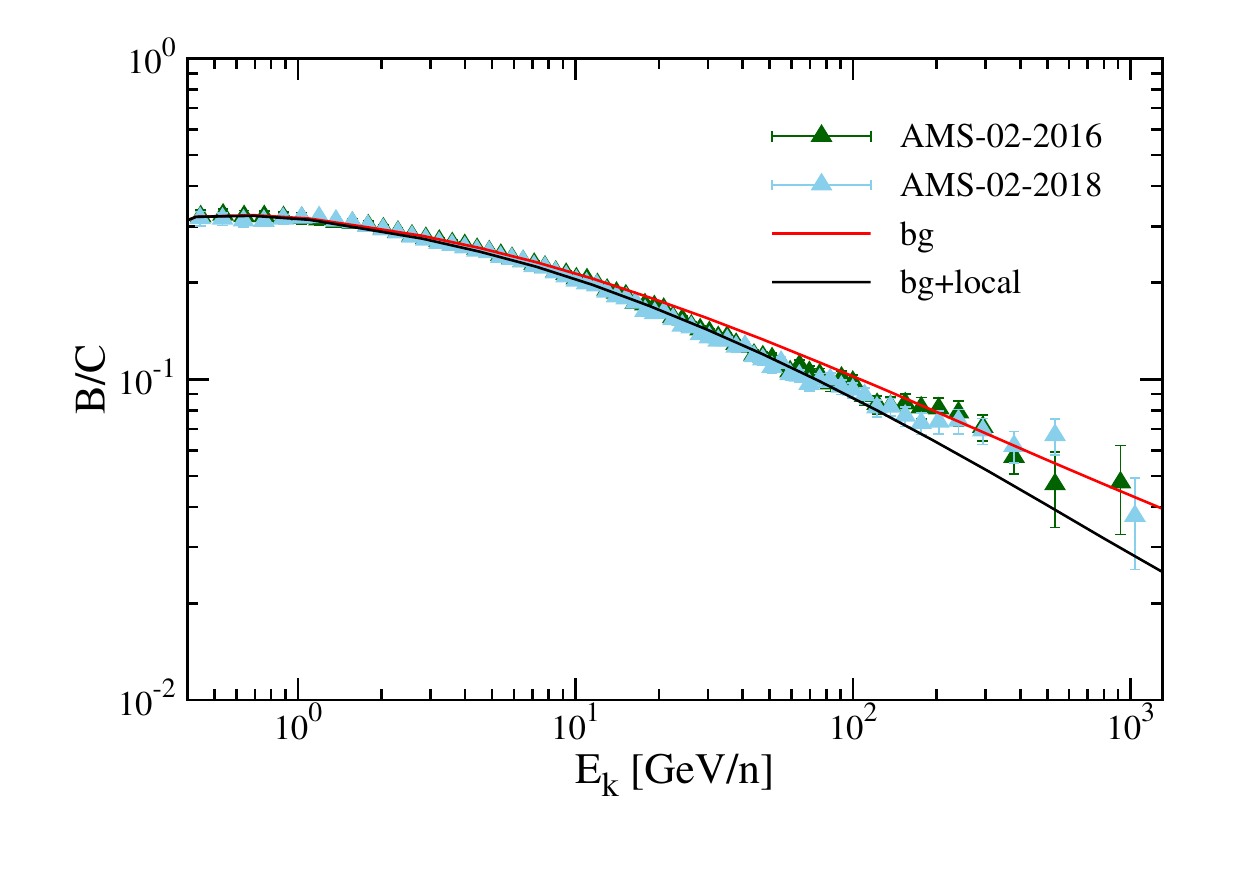}
	\includegraphics[width=0.45\textwidth]{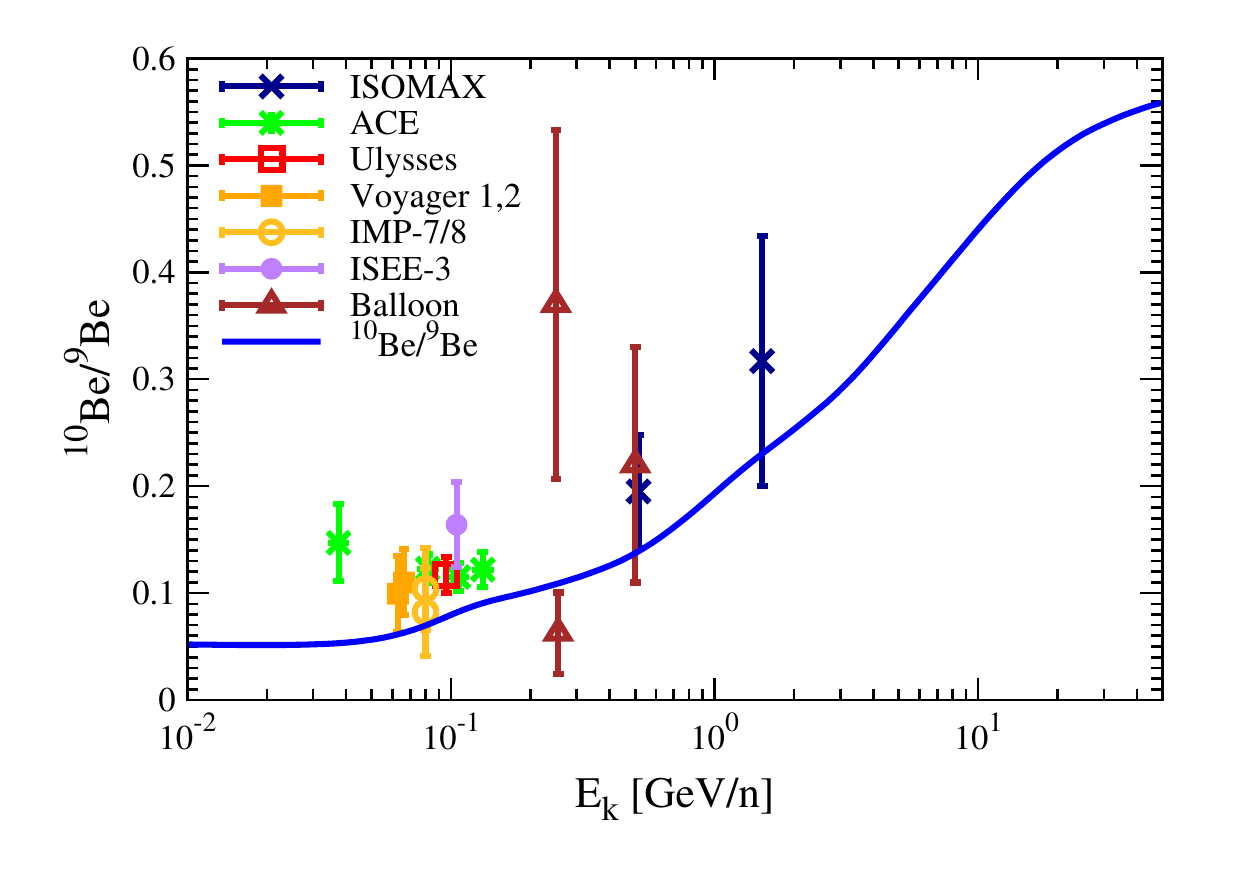}	
	\caption{
	Fit of the B/C and ${}^{10}$Be/${}^9$Be ratios to obtain the values of the SDP propagation parameters. The B/C data are taken from AMS-02 \citep{2016PhRvL.117w1102A, 2018PhRvL.120b1101A}, and ${}^{10}$Be/${}^9$Be data are taken from ACE(\cite{Yanasak_2001_ApJ}), Balloon(\cite{1977ApJ...212..262H}, \cite{1978ApJ...226..355B}, \cite{1979ICRC....1..389W}), ISOMAX(\cite{2004ApJ...611..892H}), Ulysses(\cite{1998ApJ...501L..59C}), ISEE-3(\cite{1988SSRv...46..205S}), IMP7/8(\cite{1981ICRC....2...72G}), Voyager(\cite{1999ICRC....3...41L}), respectively.
	}
	\label{fig:ratio}
\end{figure*}

The propagation and injection parameters are evaluated independently and the former can be determined by fitting B/C and ${}^{10}$Be/${}^9$Be ratios. As for the SDP model, the unknown propagation parameters are $D_0$, $\delta_0$, $N_m$, $\xi$, $n$, $v_A$ and $z_h$ respectively. Fig. \ref{fig:ratio} shows the comparison of the B/C and ${}^{10}$Be/${}^9$Be ratios between the SDP predictions and the data. The values of corresponding propagation parameters are listed in table \ref{table1}. The red lines is the B/C ratio computed only from background sources, and the black line is the one with additional carbon contribution from local source. As can be seen that, the carbon flux from the local source lowers the total B/C ratio above $\sim 10$ GeV. Within the uncertainty of the measurements, the computed B/C ratio is still consistent with the latest AMS-02 measurement. Due to lack of the precise observation, the measurements of ${}^{10}$Be/${}^9$Be have large errors, and our fitting could also account for the current data.

\begin{table}[htbp]
\caption{\textbf{Fitted spatial-dependent propagation parameters }}
\centering
\label{table1}
\begin{tabular}{ccccccc}%
\toprule
\hline
{$D_0$}& \multirow{1}{*}{$\delta_0$}& \multirow{1}{*}{$Nm$} &\multirow{1}{*}{$\xi$}&\multirow{1}{*}{$n$}&\multirow{1}{*}{$v_A$}&\multirow{1}{*}{$z_h$}\\
{$[cm^2\cdot s^{-1}]$}&&&&&{$[km\cdot s^{-1}]$}&{$[kpc]$}\\
\midrule
4.66&0.54&0.62&0.1&4&6&5 \\
\hline
\bottomrule
\end{tabular}
\end{table}



When the propagation parameters are fixed, we further study injection parameters as well as local source's age and distance. The MultiNest package is applied to perform the Bayesian inference of the corresponding parameters to obtain their posterior distributions and correlations between them allowed by the observations. The background and local source parameter set is $\vec{\Theta} = \{ A^{\rm P}, \gamma^{\rm P}, A^{\rm He}, \gamma^{\rm He}, q^{\rm P}_{0}, \alpha_{\rm P}, q^{\rm He}_{0}, \alpha_{\rm He}, {\cal R}_c, r, t, gl, gb \}$. $A^{\rm P/He}, \gamma^{\rm P/He}$ are the normalization and power index of background proton/helium flux, and $q^{\rm P/He}_{0}, \alpha^{\rm P/He}$ are the local  protons/heliums at $1$ GV and the power index. ${\cal R}_c$ is the cut-off rigidity of local CRs. $r$, $t$, $gl$ and $gb$ denotes the local source's distance, age, longitude and latitude in the Galactic coordinate system. The data include proton and helium spectra, and the dipole anisotropy amplitude. Subject to the large systematic uncertainties  of the ground-based experiments and inconsistencies in measurements, we just consider the dipole anisotropy data of AS$\gamma$ \cite{} and ARGO \cite{} experiments for the fitting.

\begin{figure*}
	\centering
	\includegraphics[height=18.cm, angle=0]{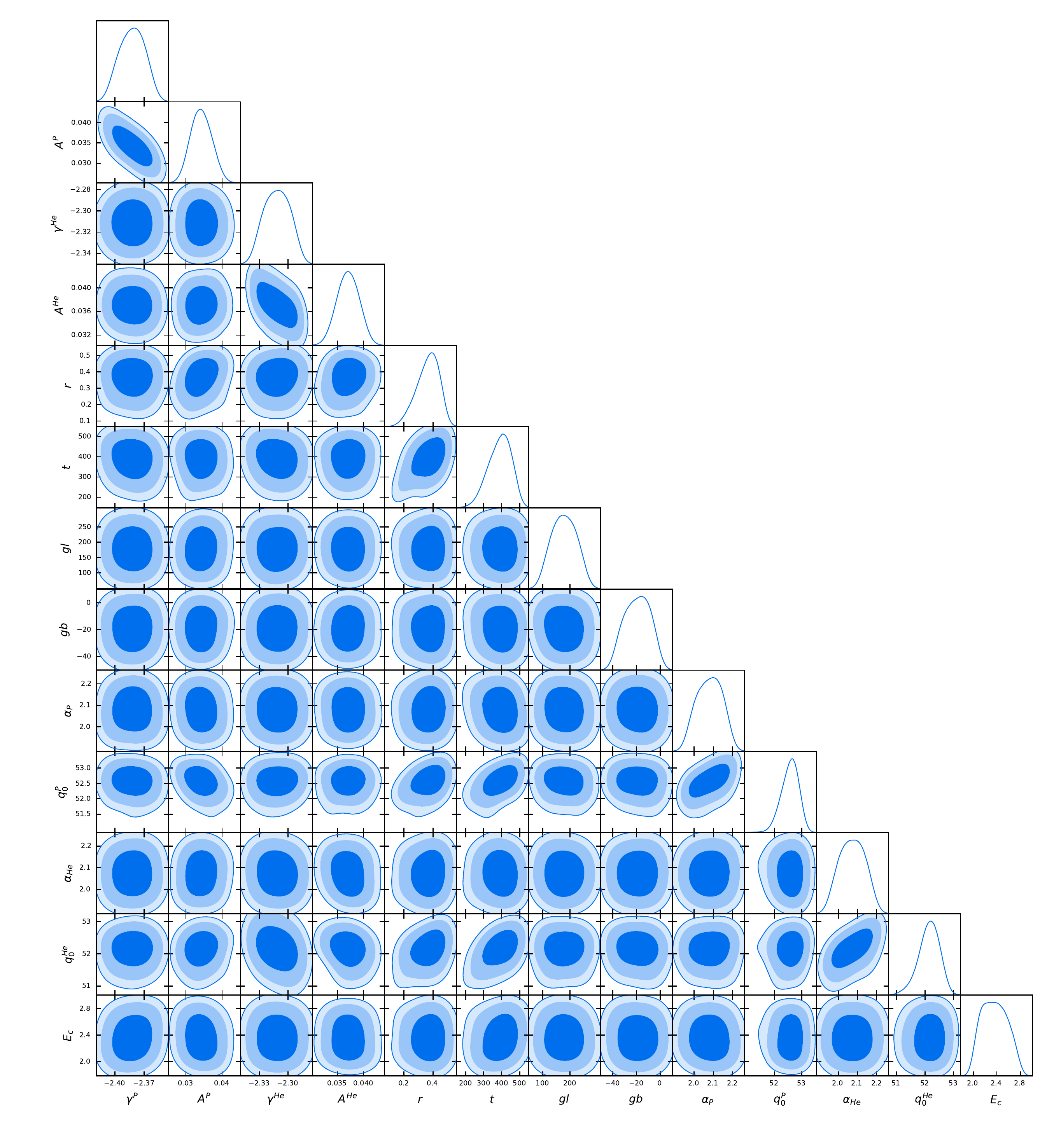}
	\caption{
		2-dimensional correlation distributions of injection parameters as well as local source's age and distance.
	}
	\label{fig:post_dist}
\end{figure*}

The 2-dimensional correlation distributions of parameters are illustrated in the triangular plot of Fig. \ref{fig:post_dist}. We also show the marginalized posterior PDFs in the diagonal regions. The dark, intermediate and light blue lines correspond to $1-\sigma$, $2-\sigma$ and $3-\sigma$ contours respectively. The posterior distributions of each parameter are listed in table \ref{table2}. For the background parameters, $A^{\rm P}$ and $\gamma_{\rm P}$ (or $A^{\rm He}$, $\gamma_{\rm He}$) are distinctly anti-correlated. This can be understood that since the injection spectrum is softer, the calculated flux at normalization energy $100$ GeV is lower and a larger normalization flux is needed in order to fit the spectrum. So do $q^{\rm P}_{0}$ and $\alpha_{\rm P}$ (or $q^{\rm He}_{0}$ and $\alpha_{\rm He}$) of the local source.

Meanwhile as can be noticed that the age and distance of the local source has strong positive correlation. For a distant local source, its age is necessary to be old due to the longer propagation distance. Otherwise, the CRs have not propagated to the solar system so far if the source is too young. Correspondingly, its injection power has to be
enhanced when its distance is far away. Therefore, $q^{\rm P}_0$ and $q^{\rm He}_0$ are positively correlated with source's distance $r$.

we also found that to explain the proton and helium spectra, the injection power index of the local source is slightly harder than the background. For example, the power index of local protons $\alpha$ is between $-2.2$ and $-2.0$, whereas it is between $-2.40$ and $-2.37$ for the backgrond. So does for helium. This has been noticed in our previous works \cite{}. Furthermore, to fit both energy spectra and anisotropy amplitude, the constraint of local source's cutoff rigidity is very tight, which is between $20$ and $28$ TeV. It seems to have no significantly correlations with other parameters, even does not change with local source's age and distance.

\begin{table}[htbp]
\caption{Fitted injection parameters of the background and local sources}
\centering
\label{table2}
\begin{tabular} {   c | l     c | l c}
\toprule
\hline
\multicolumn{2}{c}{Parameter}  &  \multicolumn{2}{c}{68\% limits}\\
\hline
\multirow{4}{*}{Background}
&{$\gamma^{P}                 $} & &$-2.382\pm 0.013             $ \\
\cline{2-4}
&{${A^{P}}^\dagger$} & &$0.0342\pm 0.0029          $ \\
\cline{2-4}
&{$\gamma^{He}               $} & &$-2.311\pm 0.013             $ \\
\cline{2-4}
&{$A^{He}                          $} & &$0.0371\pm 0.0021          $ \\
\hline
\multirow{9}{*}{Local source}
 &{$\alpha^{P}                   $} & &$2.080\pm 0.066                  $\\
\cline{2-4}
 &{$\log_{10}(q_0^P)       $} & &$52.57^{+0.36}_{-0.28}        $\\
\cline{2-4}
&{$\alpha^{He}                  $} & &$2.073\pm 0.066                  $\\
\cline{2-4}
&{$\log_{10}(q_0^{He})    $} & &$52.15^{+0.40}_{-0.32}        $\\
\cline{2-4}
&{$R_c$ [TV]} & &$23.6^{+2.1}_{-2.6}          $\\
\cline{2-4}
&{$r$ [kpc]} & & $0.366^{+0.090}_{-0.063}   $\\
\cline{2-4}
 &{$t$ [kyr] } & &$390^{+70}_{-60}                 $\\
\cline{2-4}
&gl [$^\circ$] & &$178\pm 50                          $\\
\cline{2-4}
& gb [$^\circ$] & &$-19\pm 11                           $\\
\hline
\multicolumn{2}{c}{$\chi^2/{dof}                   $} & &$158.6/86   $\\
\hline
\bottomrule
\end{tabular}
\\
$^\dagger${unit of normalization $A^{P/He}$ and $q_0^{P/He}$ are GeV$^{-1}$ m$^{-1}$ s$^{-1}$ sr$^{-1}$ and GeV$^{-1}$ s$^{-1}$ respectively.}
\end{table}

\begin{figure*}
	\centering
	\includegraphics[height=9.cm, angle=0]{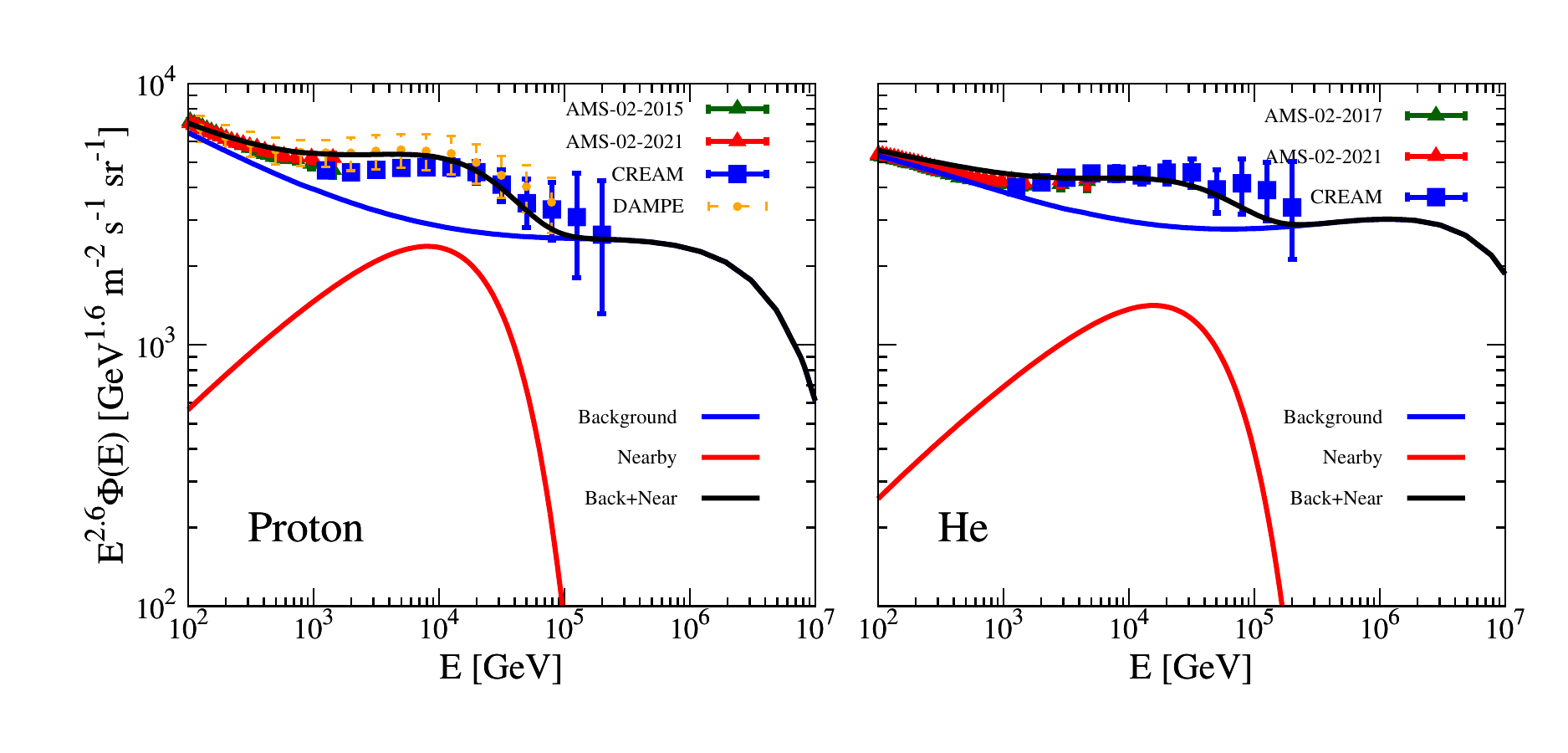}
	\caption{
Proton and helium spectra calculated with best-fit parameters. The data points are taken from AMS-02(\cite{2015PhRvL.114q1103A, 2017PhRvL.119y1101A, AGUILAR20211}), CREAM(\cite{2017ApJ...839....5Y}), and DAMPE(\cite{2019SciA....5.3793A}).
	}
	\label{fig:spec}
\end{figure*}

\begin{figure*}
	\centering
	\includegraphics[height=10.cm, angle=0]{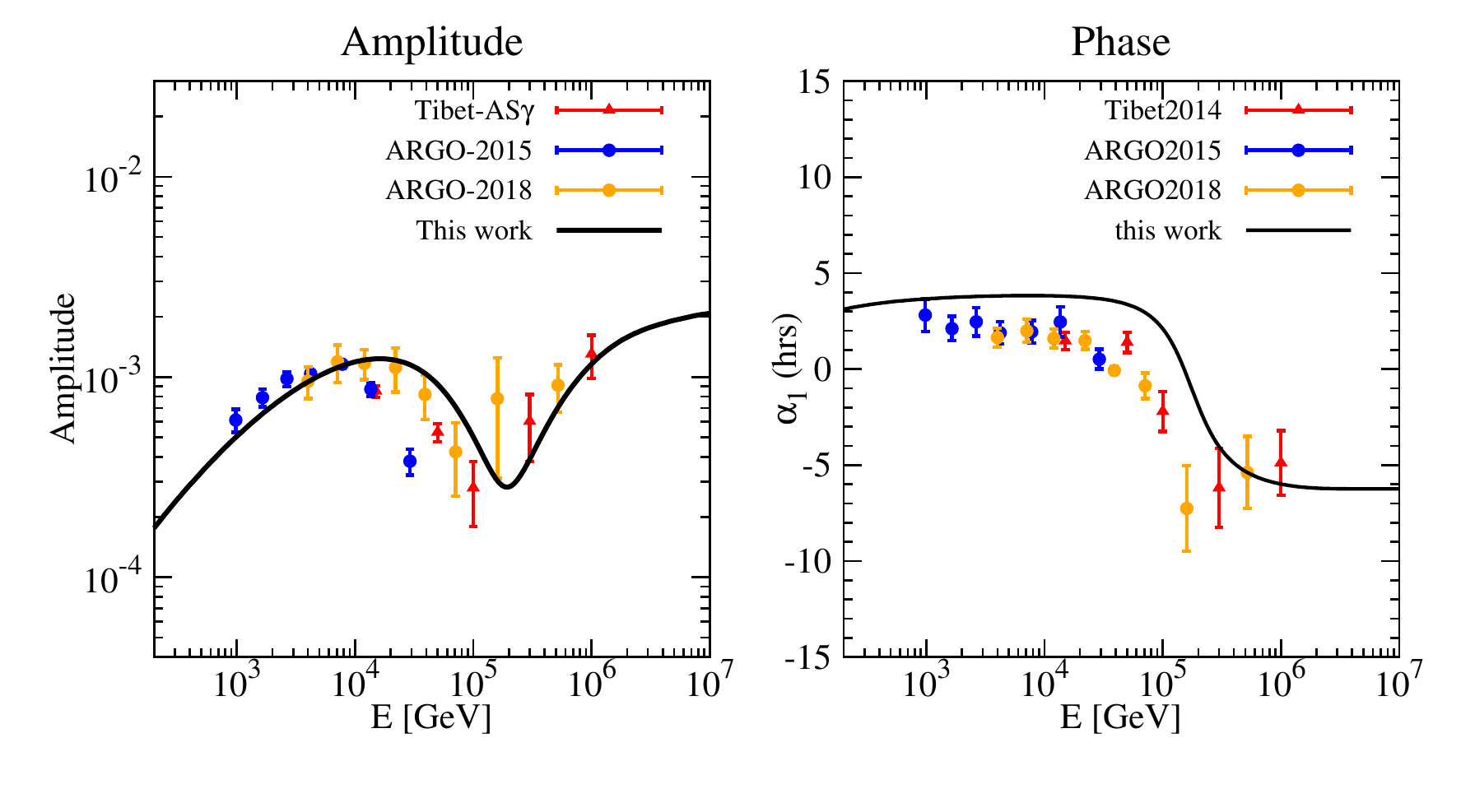}
	\caption{
Amplitude and phase of anisotropy calculated with best-fit parameters. The data points are taken from ARGO-YBJ(\cite{2015ApJ...809...90B},\cite{Bartoli_2018}),Tibet(\cite{2017ApJ...836..153A}).
	}
	\label{fig:aniso}
\end{figure*}

Fig. \ref{fig:spec} shows the calculated proton and helium spectra with best-fit parameters. And Fig. \ref{fig:aniso} illustrate the corresponding amplitude and phase of anisotropy. As can be seen that, the transition from local-source dominated to background dominated in the amplitude map is a little over $100$ TeV. This is due to large measurement uncertainties of ARGO-2018 at that energy, so that the transition does not exactly match with AS$\gamma$ data at $\sim 100$ TeV. We hope future experiments could make precise measurements of that transition.

Since we do not consider the phase of anisotropy, the calculated phase does not conform with observations less than $100$ TeV. And thus the source's position, i.e. $gb$ and $gl$, are poorly constrained, as can been seen in the Fig. \ref{fig:post_dist}. Above that energy, the CRs are dominated by the background and the phase points toward Galactic center, which is consistent with observations.

\begin{figure}
	\centering
	\includegraphics[height=9.cm, angle=0]{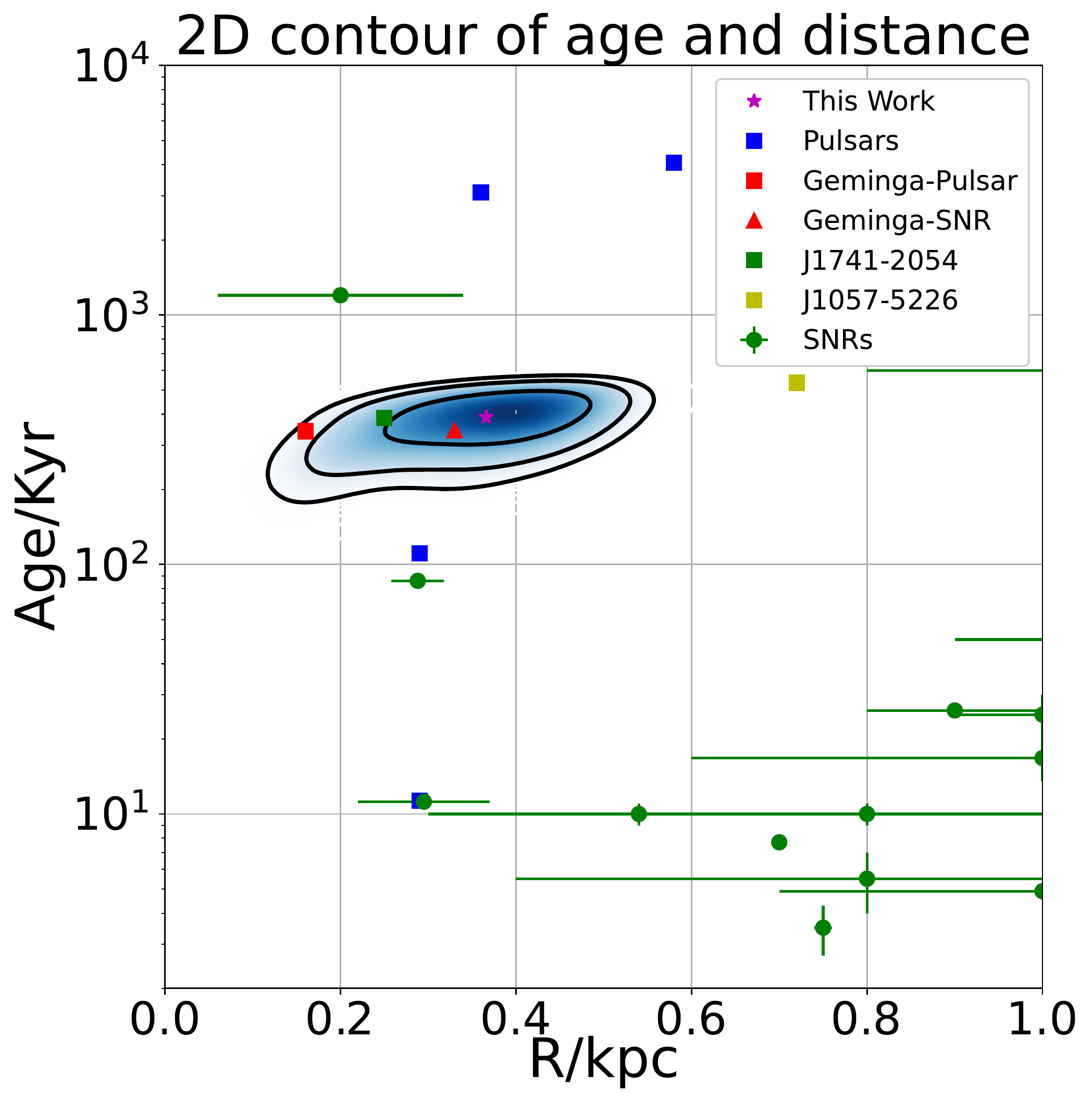}
	\caption{
2D contour of age and distance of local source. The violet star is best fit values of galactic longitude and latitude, and the solid lines are $1-\sigma$, $2-\sigma$ and $3-\sigma$ contours respectively.
	}
	\label{fig:age_dist}
\end{figure}

When the local source's position and age are constrained, we further attempt to find out the best candidate in the local catalog. Fig. \ref{fig:age_dist} shows 2-D contours of age and distance of local source with local SNRs and puslars shown as green circle and blue square points. The Geminga pulsar, J1741-2054 and J1057-5226 are shown as red,
green and yellow squares. Besides we also include Geminga SNR, which is taken as the local source to explain both energy spectra and anisoropy in our previous works. The violet star is best fit values of age and distance, and the solid lines are $1-\sigma$, $2-\sigma$ and $3-\sigma$ contours respectively. As can be seen from the figure, most of local sources are far from the contours and thus excluded. Only Geminga-pulsar/SNR and J1741-2054 are within the $3-\sigma$ regions. The Geminga-pulsar is just at $3-\sigma$ contour and Geminga-SNR are closest to the best-fit value.

\begin{figure}
	\centering

	\includegraphics[height=9.cm, angle=0]{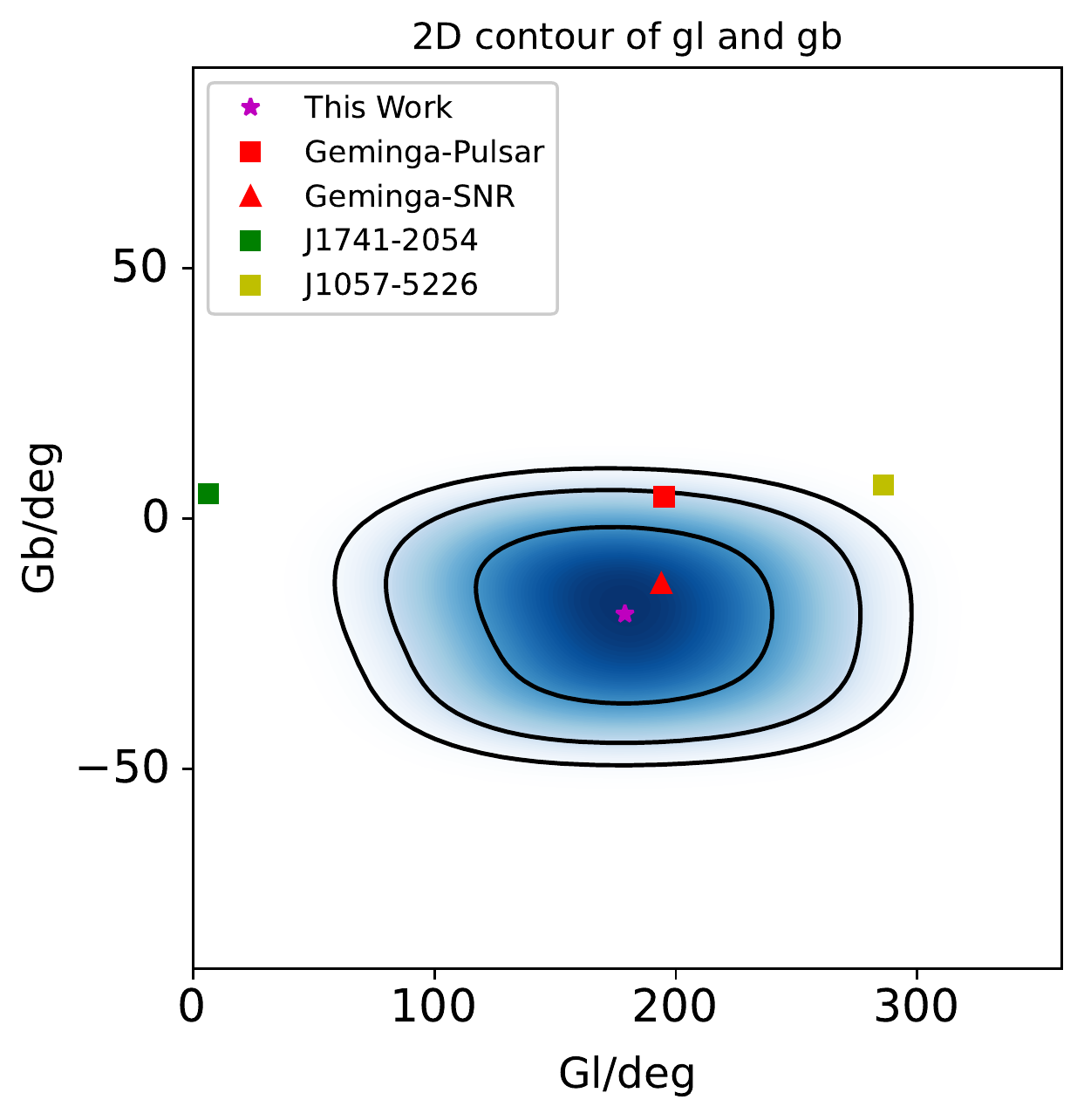}
	
	\caption{
	2D contour of galactic longitude and latitude. The violet star is best fit values of galactic longitude and latitude, and the solid lines are $1-\sigma$, $2-\sigma$ and $3-\sigma$ contours respectively.
	}
	\label{fig:gl_gb}
\end{figure}

Fig. \ref{fig:gl_gb} shows 2-D contour of galactic longitude and latitude. Due to lack of fitting phase of anisotropy, the permissible range of gl is quite large. Nevertheless, we can find that the constraints are still very tight and the allowed sources are only Geminga SNR and Geminga pulsar. The local sources at anti-Galactic center direction are very few, most of them are at the direction of Galactic center. And Geminga SNR is very close to the best-fit value. Therefore we think Geminga SNR is likewise the probable candidate of the local source.

\section{Conclusion}

We have built up an unified scenario, based on SDP+local source, to explain the observations of both energy spectra and anisotropy of CR nuclei below PeV energies. We find that less than $\sim 100$ TeV, not only does the
local source contribute to the spectral hardenging at $\sim 200$ GV and subsequent softening at $20$ TeV in the energy spectra, but also dominates the GCR streaming and determines the low energy anisotropy
pattern. From the phase of the dipole anisotropy, we infer that the SNR associated with
Geminga may be an important candidate source.

In this work, we further investigate the injection parameters and local source's position and age in detail by the aid of the Bayesian inference tool, MULTINEST. We find that the age and distance of the local source are positively correlated. For a distant local source, its age has to be older. Otherwise, the CRs could not propagate to the solar system currently if the source is too young. And the corresponding injection power is enhanced, so $q^{\rm P}_0$ and $q^{\rm He}_0$ are also expected to be positively correlated with source's distance $r$. Meanwhile both energy spectra and anisotropy amplitude severely constrain the local source's cutoff rigidity, which is between $20$ and $28$ TV.

The energy spectra and anisotropy amplitude also give strong constraints of local source age and distance to the solar system. Most of local sources have been excluded and we find that only Geminga SNR is very close to the best-fit value. In addition, despite that the permissible range of $gl$ and $gb$ is quite large, most of local sources are also excluded, since most of them are in the direction of Galactic center. Only Geminga SNR is very close to the best-fit value. From these, we infer that the Geminga SNR could be the probable candidate of local source.

Since the Geminga SNR is still away from the direction of observation phase, we do not include the amplitude of anisotropy into fitting. In the future work, we would take the influence of local ordered magnetic field into account to better reproduce the observational phase less than $100$ TeV. And based on the reproduced phase, we would like to find out the candidate of CR nuclei source from the local catalog.

\section*{Acknowledgements}
This work is supported by the National Key Research and Development Program of China (No. 2018YFA0404202), the National Natural Science Foundation of China (Nos. 11635011, 11875264, U1738209, U2031101, U2031110).

Software: DRAGON (\cite{2008JCAP...10..018E, 2017JCAP...02..015E}) available at https://github.com/cosmicrays.

MULTINEST (\cite{2008MNRAS.384..449F, 2009MNRAS.398.1601F}) available at https://github.com/farhanferoz/MultiNest.

\bibliographystyle{apj}
\bibliography{ref}

\end{document}